\documentclass[10pt,twocolumn]{article} 
\usepackage{simpleConference}
\usepackage{times}
\usepackage{graphicx}
\usepackage{amssymb}
\usepackage[hyphens]{url}
\usepackage{fancyvrb}
\usepackage{hyperref}

\usepackage[tight,footnotesize]{subfigure}

\usepackage{tikz}
\newcommand\encircle[1]{%
	\tikz[baseline=(X.base)] 
	\node (X) [draw, shape=circle, inner sep=0] {\strut #1};}

\begin{document}

\title{An Historical Analysis of the SEAndroid Policy Evolution}

\author{Bumjin Im \quad\quad Ang Chen \quad\quad  Dan Wallach \\
Rice University, Houston, Texas, USA\\
\today\\
\{bi1, angchen, dwallach\} @rice.edu
}

\maketitle
\thispagestyle{empty}

\begin{abstract}

Android adopted SELinux's mandatory access control (MAC) mechanisms in
2013. Since then, billions of Android devices have benefited from
mandatory access control security policies. These policies are expressed
in a variety of rules, maintained by Google and extended by Android OEMs.
Over the years, the rules have grown to be quite complex, making it
challenging to properly understand or configure these policies.

In this paper, we perform a measurement study on the SEAndroid repository
to understand the evolution of these policies. We propose a new metric to
measure the complexity of the policy by expanding policy rules, with their
abstraction features such as macros and groups, into primitive ``boxes'',
which we then use to show that the complexity of the SEAndroid policies
has been growing exponentially over time.  By analyzing the Git commits,
snapshot by snapshot, we are also able to analyze the ``age'' of policy
rules, the trend of changes, and the contributor composition. We also look
at hallmark events in Android's history, such as the ``Stagefright''
vulnerability in Android's media facilities, pointing out how these events
led to changes in the MAC policies. The growing complexity of Android's
mandatory policies suggests that we will eventually hit the limits of our
ability to understand these policies, requiring new tools and techniques.
\end{abstract}

\section{Introduction}

Smartphones are a primary target of malicious attacks~\cite{la2013survey}. The Android system---as it holds the largest share of the mobile OS 
market---has unfortunately become a prominent attack target as well~\cite{zhou2012dissecting}. 
Over the years, many attacks have been reported, such as Stagefright~\cite{stagefright}, Blueborne~\cite{blueborne}, and Toaster~\cite{toaster}, 
each of which has led to significant security concern. 
In order to harden Android against such threats, researchers and developers
have adapted a wide variety of security mechanisms to the Android
environment, including process separation, finer grained access control,
and secure booting~/ remote attestation. 
This study focuses on one such mechanism: SEAndroid~\cite{smalley2013security}. 

As an extension of SELinux~\cite{peter2001integrating}, SEAndroid was originally introduced by the NSA in 2013.
It performs mandatory access control (MAC) to enforce security policies, regulating whether a particular subject 
(e.g., a process) can perform a certain action (e.g., read/write) on an object (e.g., a file/socket). 
It achieves this by referring to a pre-installed security policy with a set of access control rules, which are 
compiled into a database and loaded by the kernel at boot time.

As with SELinux, configuring a SEAndroid policy is not an easy task~\cite{jaeger2003analyzing,schreuders2012towards}. It is often far from obvious to reason about 
whether a particular set of rules achieve a desired policy, or even to understand what policy certain rules try to implement. 
There are several reasons for this. First, the policy language allows many abstraction features, such as groups, attributions, and (nested) macros, 
which make it challenging to infer the scope of individual rules. Second, the policy rules have little accompanying documentation. 
Moreover, they evolve significantly every year.

Fortunately, Google's default SEAndroid policy is maintained in a Git repository~\cite{sepolicy_repo} as part of the Android Open Source Project (AOSP)~\cite{aosp}. 
The Git history provides us with a detailed chronicle of all changes to the policy rules, as well as 
the commit messages associated with each change, totaling more than 13,000
commits over SEAndroid's history. These commits serve as 
a rich source of information for us to perform an historical analysis of the policy's evolution, snapshot by snapshot. 

Performing this analysis involves at least two challenges. First, we need a good metric to quantify the complexity of a particular 
policy snapshot. Simply counting the \textit{number} of rules in a snapshot is not enough, because the SEAndroid policy language 
supports a variety of abstractions and grouping concepts, making a rule count less useful. 
We address this challenge by designing a new metric, \textit{boxes}. This metric views all possible access control policy rules as forming a 
four-dimension space, with the axes being \textit{subject}, \textit{object}, \textit{class}, and \textit{permission}. 
Each point in this space is called a \textit{box}, representing a smallest ``unit'' in the rule space. 
By analyzing how many rules touch the same box, and how many boxes are impacted by any given
rule, we can then quantify the policy's complexity. Consider that complex rules may touch 
many boxes and many of those boxes may be impacted simultaneously by multiple rules.
This means that a change to any one rule may or may not result in a change to the resulting security policy!
If an engineer truly desires to change a given box, it becomes necessary to go hunting for every
possible rule that might overlap with it.

Our second challenge is that there are thousands of Git commits that impact
the SEAndroid policies.
We use a combination of approaches to address this. We have designed and
implemented an automated system that collects and analyzes each
Git commit, identifying ``jump points'' in complexity both in terms of the number of rules and the number of boxes. 
For significant jump points, we also manually inspect the commit messages associated with these commits, and perform a differential analysis on the rules 
before and after each such commit to understand the rationale of the changes. Furthermore, we use the timestamp information to ``match'' the commits with 
historical events of Android security, and analyze how these events are reflected in the policy's evolution. 

Using the above metrics and methodology, we have performed an historical analysis on the SEAndroid policy's evolution. 
We focus on the development of security metrics that we can derive from these policies, such as the number of boxes and rules, 
the evolving list of types and macros, the different authors contributing to the policy over time, how SEAndroid 
policy stabilized over time, as well as the hallmark events in Android history. 
Based on these measurement results, we also provide insights into how SEAndroid might 
evolve to become simpler and more useful. 

The structure of the paper is as follows. After describing more background 
material in Section~\ref{sec:background}, we introduce our measurement methodology in Section~\ref{sec:methodology}, present measurement results 
in Sections~\ref{sec:measurement} and \ref{sec:historical}. Then, we discuss several related topics in Section~\ref{sec:discussion}, 
present related work in Section~\ref{sec:related}, and conclude in Section~\ref{sec:conclusion}. 
\section{Background}
\label{sec:background}

In this section, we present more background material on Android security architecture and, in particular, SEAndroid. 

\subsection{Android security architecture} 

\begin{figure}[h!]
	\centering
	\includegraphics[width=6cm]{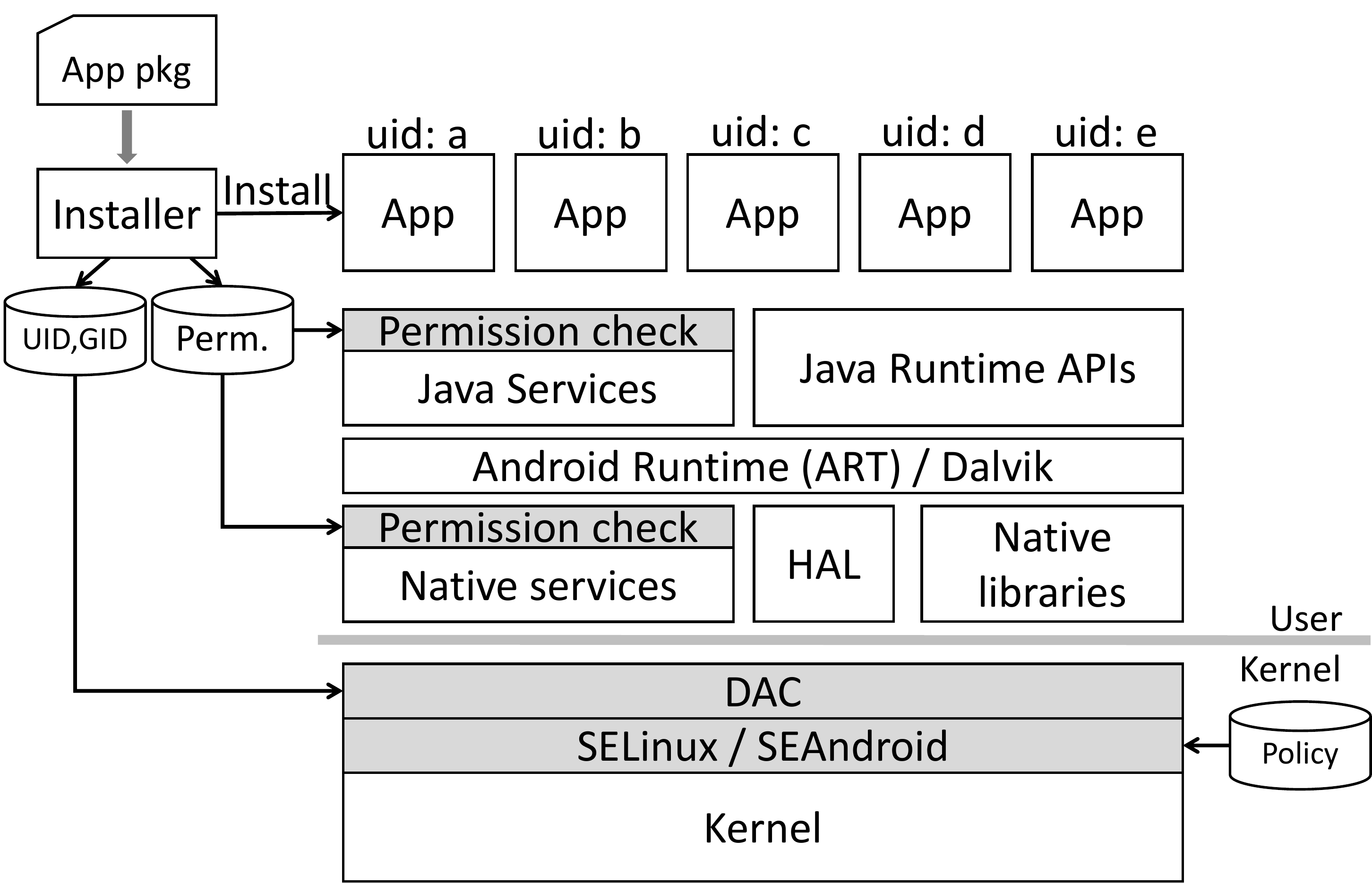}
	\caption{The Android security architecture.}
	\label{architecture}
\end{figure}

Figure~\ref{architecture} shows the multiple layers of security mechanisms Android uses to protect system resources and user data~\cite{elenkov2014android}. 

\textbf{Install-time permissions.}  Every Android application includes a
``manifest'' file specifying the app's desired permissions. Prior to
Android~6.0, the user was queried at install-time whether the desired
permissions were acceptable (i.e., all-or-nothing). After this, the new app
is assigned a distinct Unix ``user'' ID, allowing traditional Unix-style
file permissions to separate the storage for each app. However, Android
includes a variety of system services, speaking over {\em Binder} (an
interprocess communication channel mediated by the OS kernel). Each Binder
service is responsible for determining if its caller is permitted to use
it, querying a central database constructed from those install-time
permissions. This resulted in permission checks occurring all over the
Android software base, in both Java and native code. This, in turn, made it
a challenging research project just to produce a mapping from every Android
API call to its corresponding set of required permissions~\cite{demystified}.

\textbf{Time-of-use permissions.} 
As of Android~6.0, Android apps still declare their permission requirements
in the manifest, but those permissions are not granted until the app
explicitly requests them from the user, preferably at time-of-use. Users are
free to deny permissions and even to revoke permissions later on from the
system settings. The underlying enforcement of these permissions is largely
the same as before, with individual Android services querying whether their
caller has a necessary permission.

\textbf{Classic Unix discretionary (DAC) permissions.}
Of course, Android is also just another flavor of Unix. If an application
directly accesses a Unix resource such as a file or device, traditional Unix
user and group IDs manage the security. A Unix group ID is preassigned to each
permission, and the Android application launcher assigns all the necessary
group IDs to the application process when it is launched. Since a Linux
process can possesses multiple group IDs, the access will be granted or
denied appropriately.

\textbf{SEAndroid mandatory (MAC) permissions.}  
After the user ID and group ID checks are performed,
SEAndroid~\cite{smalley2013security}, which is an extension of
SELinux~\cite{peter2001integrating} for Android, can additionally check all
system calls against its own policy. This policy is loaded at boot time and 
enforced inside the OS kernel. SEAndroid policies, by their static nature, 
cannot be changed at runtime to reflect new applications and user-expressed
permissions.  They can, however, be used to isolate system services, file
directories, and even Binder resources.

\subsection{Example: Location services}

\begin{figure}
	\centering
	\includegraphics[width=6cm]{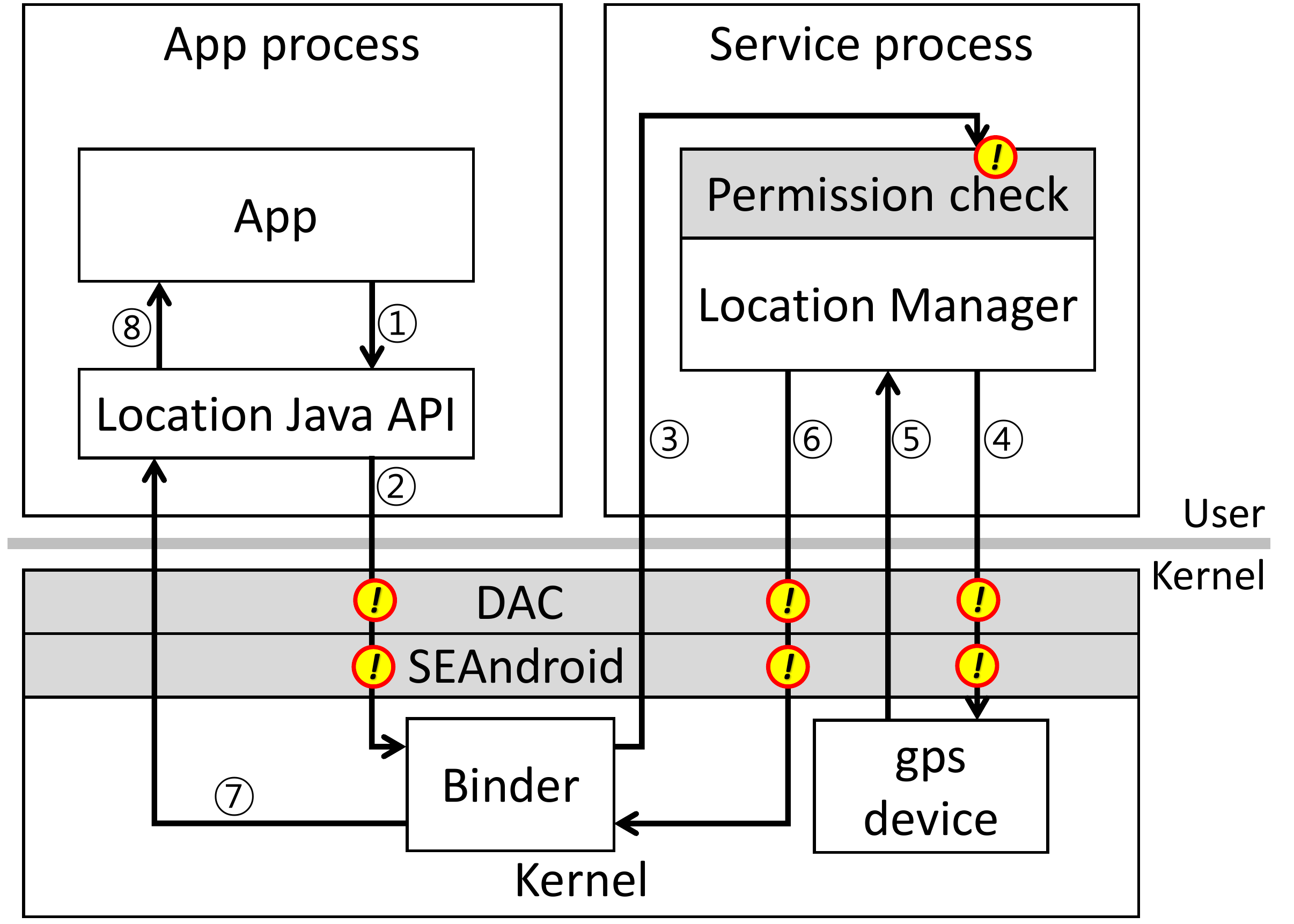}
	\caption{Control flow with the Android location API. Seven
          permission checks are highlighted with yellow circles.
	\label{callflow}}
\end{figure}

To show how these different security systems interact, Figure~\ref{callflow} diagrams the control flow involved when an app makes a call to the Location Services, which manages the GPS device and other location-related (and thus, privacy-sensitive) features.

If an application calls a location API \encircle{1}, the corresponding
library establishes a Binder channel between the application and the
\textit{location manager} \encircle{2}. Since Binder is an IPC mechanism
provided by the kernel, the client requires a corresponding discretionary
access control (DAC) permission which is always allowed. After that, a
SELinux hook is called to check the SEAndroid permission, which is also
configured to always permit this call. Binder transfers the request to the
location manager \encircle{3}, which then calls a {\em checkPermission()}
system API (shown as ``permission check'') which checks if the user granted
the relevant location permissions to the calling app.
If granted, the location manager interacts
directly with the GPS device \encircle{4}. However, this access also
requires DAC and SEAndroid enforcement, which are configured so no other
subject in the Android universe is permitted to interact directly with the
device. The API response unwinds the call path\encircle{5} -
\encircle{8}. Each step includes another opportunity for DAC and
SEAndroid permissions to be checked, but these are all permitted by
default.

Even in this simple example, there are seven different access control
enforcement opportunities on the path from the app to the GPS device
driver. In this case, DAC and SEAndroid only serve to ensure that the 
GPS device is only reachable from the location manager service, while the
permission checking for the app is handled internally by the location
manager. As this example illustrates, Android security enforcement is
complex to analyze.

Here, the value of SEAndroid is to protect system resources more
precisely from low-level attacks. 
But as we see in Figures~\ref{architecture} and~\ref{callflow}, the
SEAndroid policy does not have any relationship to the Android security
model as Android users and app programmers understand it. 
In addition, there is no formal documentation specifying any of the
SEAndroid policies, much less how they interact and how the user-visible
permissions or Unix discretionary permissions are meant to interact with
each other and with SEAndroid.

\subsection{SEAndroid policy rules}

The SEAndroid policy engine was present on Android devices since the early 2010s, but it was configured in an inactive ``permissive mode''. 
The policy was initially configured by NSA and committed to the AOSP repository~\cite{sepolicy_repo}. Any Android device OEM can add their own 
policy along with the original policy, e.g., using tools such as \texttt{setools}~\cite{setool} and \texttt{audit2allow}~\cite{audit2allow}. 
This is typically done by examining denial logs and adding additional rules, which can then be submitted as patches for 
inclusion in future ASOP releases. 

To enforce the policy, the rules are then parsed and compiled by a compiler called \texttt{checkpolicy} and loaded in the kernel at boot time.
Needless to say, configuring a SEAndroid policy is critical, but at the same time, not an easy task~\cite{smalley2002configuring,schreuders2012towards}. 
A misconfigured policy will lead to misbehaviors or even security vulnerabilities. In order to understand the complexity of the policy, 
we need to first understand what the rules look like. 

\textbf{Allow rules.} The majority of the policy consists of a sequence of ``allow rules'', such as 
     \texttt{allow appdomain zygote\_tmpfs:file read},  
which states that any subject of the type \texttt{appdomain} (i.e., all applications) 
should be granted access to any \texttt{file} object of the type \texttt{zygote\_tmpfs} (i.e., files that the \texttt{zygote}
 process created in the \texttt{tmp} file system). 
Besides \texttt{file}, typical \textit{classes} also include directory, socket, process, and so on, each indicating a particular category of resources. 
This rule additionally specifies a read \textit{permission}, meaning that accesses can only happen via the \texttt{read} system call. 
As we can see, such rules allow for very fine-grained policy, but writing a rule for each subject and object/class pair for every system call 
would not scale. 

\textbf{Abstraction features.}  To avoid the policy size from growing out
of control, the SEAndroid/SELinux policy language supports a number of
abstraction features, such as wildcards, groups, negation, and complements.
A wildcard ({\texttt *}) can be used in any object, class, or
permission fields to represent ``any possible entity''.  For instance, a
rule \texttt{allow appdomain zygote\_tmpfs:file *} would allow any system
call, not just \texttt{read}.  One could additionally group several
entities together, such as \texttt{allow appdomain zygote\_tmpfs: {file
    dir} *}.  The complement feature is indicated using a tilde
({\raise.17ex\hbox{$\scriptstyle\sim$}}), such as \texttt{allow init
  fs\_type:filesystem {\raise.17ex\hbox{$\scriptstyle\sim$}}relabelto},
meaning that any system call \textit{but} a \texttt{relabelto} is
permitted.

\textbf{Neverallow rules.}   ``Neverallow'' rules works as assertions for debugging. 
If such a rule overlaps with an allow rule, the policy compilation would report an error. 
These rules have a similar format with allow rules, such as \texttt{neverallow domain init:binder call}.

\subsection{The complexity of SEAndroid policy} 

Over the years, the SEAndroid policy has grown to be very complex. For
instance, the snapshot as of July 2017 specifies 91 classes and 1,603
permissions.  In addition to this, three more factors exacerbate the
complexity.

First, the use of abstraction features only hide the complexity but does
not eliminate it.  For instance, \texttt{allow untrusted\_app self:file *}
may be easier to read, but the wildcard also provides an opportunity to
sweep important issues under the table.  Similar to abstraction features,
SELinux supports DTE~\cite{badger1995practical}, which allows a domain to
be associated with multiple types.  This significantly decreases the number
of rules, but again, hides complexity and potential vulnerability.

Second, the existence of neverallow rules itself hints at the complexity of the policy. 
In fact, SEAndroid is by nature a \textit{mandatory} access control mechanism, which means that 
the default action for anything is already ``disallow'', unless explicitly granted access by an allow rule.
Providing such assertion tool for policy configuration of a mandatory access control mechanism 
implies that there is concern for debuggability and for accidentally allowing too many actions.

Third, the policy engineering practice is at times idiosyncratic.  One
common practice of developers, for instance, is to extract denial messages
from the kernel log using tools like \texttt{audit2allow} and simply create
a new policy to allow them. Security concerns aside, this tends to create
unoptimized, messy policy rules.

The combination of the above factors mean that it becomes difficult for an
analyst to read the rules and understand exactly what a given local change
might entail in the final calculus of what is allowed and what is denied.
Such complexity, in fact, is one of the motivating factors for us to
perform this analysis.

\section{Methodologies}
\label{sec:methodology}

To the best of our knowledge, we are the first to perform an historical analysis of the SEAndroid policy. 
This is in contrast to previous work, 
such as Wang et al.~\cite{wang2015easeandroid} that use machine learning to improve the policy, 
and Zanin et al.~\cite{zanin2004towards} that use formal methods to verify policy correctness. 
Our goal is to understand \textit{not only the most recent snapshot, but also how the policy 
has evolved over time}. 
Since its introduction, more than 16,000 commits have accumulated in the repository, with detailed timestamps, author information, and commit messages, 
providing a valuable source of information for understanding SEAndroid. 

\subsection{The ``box'' metric} 

One might wonder whether the complexity of a policy snapshot could simply
be measured by counting the number of rules.  However, this is not enough
due to the abstraction features, such as macros, groups, and wildcards,
that are heavily used throughout the policy base.  To address this, we have
designed a new metric, the ``box'', which is similar in spirit to Lampson's
term ``attribute'' in his original work on access control
matrices~\cite{lampson1974protection}. At a high level, a box is a
quadruple with one subject type, one object type, one class, and one
permission. This is the atomic unit that we use to quantify the
complexity. With this, we can look at two interesting metrics: how many
rules target each box, as well as how many boxes are targeted by each
rule.

\textbf{From rules to boxes.} In order to obtain the boxes, we decompose each rule in the policy base in the following way. 
First, we expand all the macros used in the policy rules, and obtain all the classes and permissions from the \texttt{access\_vectors} file. 
We then scan all the policy files and obtain all attributes and types. After this, we perform a second pass over all the policy files 
and decompose each {\tt allow} and {\tt neverallow} rule to their respective boxes, using the subject, object, class, and permission fields in each rule. 
A single allow rule may be decomposed into many {\tt allow} boxes; {\tt neverallow} rules are decomposed as ``negative'' boxes which are then subtracted from 
the {\tt allow} boxes. The final outcome is a list of {\tt allow} boxes equivalent to the original policy.

\subsection{Git repository analysis}
\label{sec:git-rebase-merge}

We repeat this analysis using boxes for each snapshot contained in the Git history. 
A Git history is a directed acyclic graph (DAG), where the commits are vertices and parent-child relations are edges. 
Because Git supports multiple branches which might then merge, not every
pair of Git commits is necessarily going to have an ancestor-descent
relationship.
By analyzing the graph along with the contents of each vertex, we would be able to understand how the 
policy evolved over time. In addition to analyzing the source code ``diff'' in each commit, a commit also has various types of metadata, 
such as the committer's timestamp, email address, and comments.
Such information allows us to gain further understanding of the rationale behind each policy change. 
Last but not least, based on the timestamps of the Git commits, we can also associate these changes to important historical events, 
such as security breaches. 

One complication that arises due to the nature of Git is the large number
of branches, as it is a \textit{distributed} source code management system
with many contributors. A contributor can create new branches any time
without communicating with the master branch, and merge these branches
back. Consequently, if we looked only at the timestamp of each commit and ordered every
commit by these timestamps, we would be looking at multiple interleaved
histories. Git also supports a ``rebase'' feature that allows separate
branches to be rearranged into single linear timeline, albeit with the
original timestamps, although multiple changes can also be ``squashed''
post-facto into a single commit event. Suffice to say that our view of the Git repository,
as it's delivered to us, certainly represents the evolution of the SEAndroid
policy over time, but it's possible that we're not seeing important parts 
of the history.

To simplify our analysis, we decided to perform our measurement study on
the master branch only. Although looking at other branches may give
additional information, we believe that an analysis of the master branch is
a useful starting point, because the master branch reflects the history of
Android as it was shipped from Google through AOSP to the Android OEMs.

\subsection{Our measurement tool} 

We build our measurement tool in Python using 2,000 lines of code, with three components. 
Our crawler uses standard Git commands to check out the repository snapshots, our parser generates the box metrics from rules, 
and our serializer uses Python's \texttt{msgpack} library to store the policy to the disk. 

On our experiment platform, an Intel Core i7 computer with 4~cores and 32~GB RAM and a 4~TB hard disk, our tool takes several seconds 
per policy snapshot, parsing all rules and generating boxes from the rules. 
For all commits from January 2012 to August 2018, our tool generates 3~TB
of raw data, with hundreds of thousands of boxes per commit. 
Constructing the full database takes more than 90~hours of processing. 
\section{Measurement Results}
\label{sec:measurement}

In this section and the next, we present the measurement results obtained from 16,100 commits to the SEAndroid policy repository between  
January 2012 and August 2018. We focus on the results obtained using the new box metric in this section, and provide a broader, historical 
analysis in the next section. 

\subsection{Boxes vs. rules}\label{chapnumbox}
\begin{figure}
	\centering
	\includegraphics[width=8.5cm]{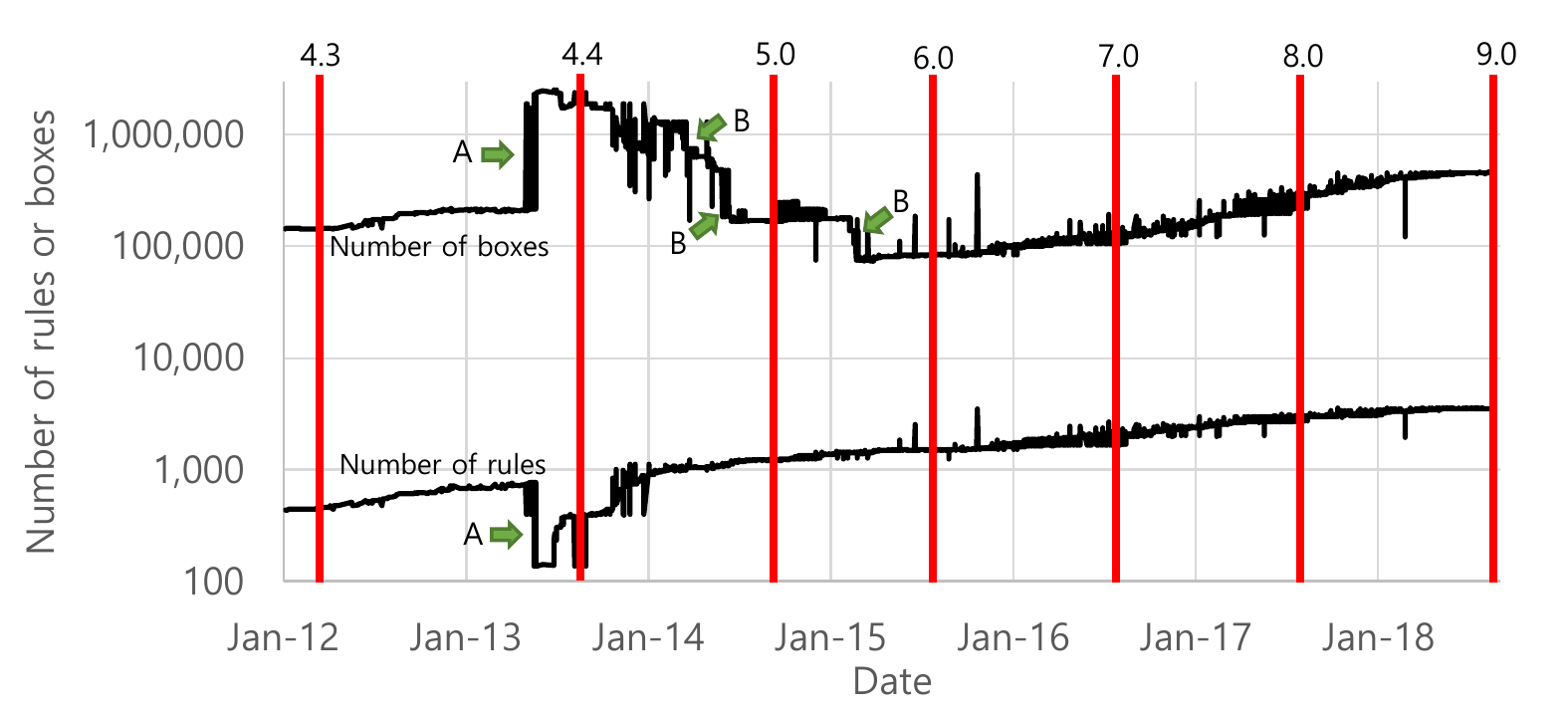}
	\caption{The number of rules vs. the number of boxes. \label{numrules}}
\end{figure}

Figure~\ref{numrules} shows two time series of the number of rules versus the number of boxes in each commit. 
We can see that, except for the period of time between mid-2013 and mid-2015, the curves are increasing 
roughly exponentially (note that the Y-axis is in logarithmic scale). 

Zooming in on the timeline between mid-2013 and mid-2015, we can see many fluctuations in both curves. 
We have in particular identified two events that we marked as ``A'' and ``B''. 
In the ``A'' commit, authored by a Google engineer, the number of rules suddenly dropped from 1,000 to 200; but the number of boxes jumped 
from 214,000 to 2,315,000, an increase of more than 10$\times$. We have manually analyzed this commit, as well as the differences from the previous commit,  
and found that this new commit associated all types to a single domain---\texttt{unconfineddomain}. 
This new domain allowed virtually any access to any object, and removed all other individual access rules with the same subjects and objects, 
which represents a clear break from past practice. Interestingly, we found that this ``unconfined domain'' 
was present even in the original version authored by NSA; but at that time, it was intended for some special entities that should 
bypass all security enforcement, and nothing was associated with it. 

So what was Google trying to accomplish with this one commit? Here is the original commit message: 

\vspace{2mm}
\hspace{3.15mm}\fbox{\begin{minipage}[t]{0.4\textwidth}
	{\tt \small{Make all domains unconfined.
		
		This prevents denials from being generated by the base policy.
		Over time, these rules will be incrementally tightened to improve
		security.}
        }
\end{minipage}}

\vspace{2mm}

\noindent We are not able to find any external documentation for this drastic
decision, but we could draw inferences about what problems they may have
been facing. In 2012, Android passed iOS in market
share~\cite{mahapatra2013android}, having roughly double the market share
of iOS in 2013. Such drastic growth and intense competition may have
created pressures to ship code on time, with security necessarily a lower
priority.  As the commit text suggests, Google had long-term plans for
improving its use of SEAndroid, but for now wasn't planning on using it
for anything beyond the benefit of having it integrated at all, allowing
OEMs to begin experimenting with SEAndroid security policies.

Indeed, Google did fix it later. The ``B'' arrows indicate changes to the
SEAndroid policy that drastically shrank the size of the ``unconfined
domain'', and the general downward slope of the number of boxes during this
time period shows a diligent effort over 1.5 years to ultimately eliminate
the unconfined domain from the SEAndroid policy. By the Android 6.0
release, the unconfined domain was no longer in use.

We note the disconnect between the number of rules and the number of
boxes. Each of the ``B'' arrows shows a significant reduction in the number
of boxes, yet there is no corresponding change in the number of rules. This
suggests an engineering process of methodically adding focused rules to
cover the needs of various applications that were previously satisfied by
the unconfined domain. After policy testing, large chunks of the unconfined
domain could be unnecessary and therefore removed. We see this in several
large downward steps in the number of boxes, as well as in the
broader downward slope during this time period.

As of 2018, Android is even bigger than before, there are more participants than before, and the policy is no more changing rapidly, so we do not expect that a similar event is happening, but this is a great example for us to suggest we need proper metrics on the policy configuration as well as the documentation.

\vspace{1mm}\noindent\textit{Takeaway \#1:} 
Such results illustrate the importance of considering rules and boxes as distinct metrics of policy complexity. Even though in more recent years we see the rules and boxes growing side by side, there is demonstrably no necessity that there be a linear relationship between rules and boxes.
\subsection{Number of boxes in a rule}\label{subsecbpr}
\begin{figure}
	\centering
	\includegraphics[width=8.5cm]{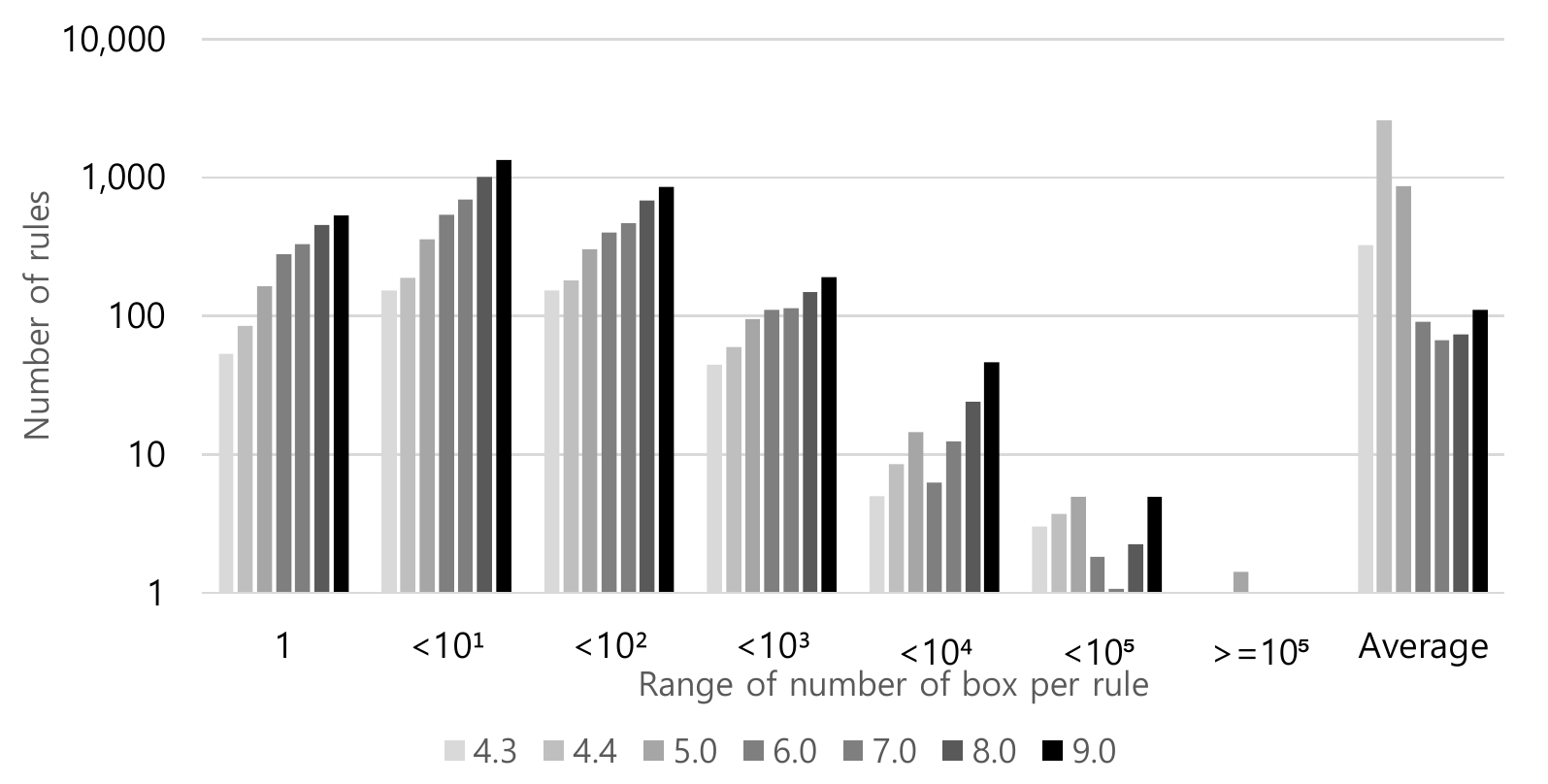}
	\caption{Average number of boxes per rule.\label{avgnumbyrule}}
\end{figure}

As we noted earlier, the number of boxes or rules, by themselves, do not necessarily tell a complete story about the complexity of a given security policy. We next look at the number of boxes per rule, which measures the complexity due to the use of macros and other grouping features of SELinux security policies.

Figure~\ref{avgnumbyrule} shows these ratios as of each major Android
release from Android~4.3 through 9.0. We then group the results into
frequency buckets on a logarithmic scale, so ``$<10^1$'' counts the number
of rules that touch $2-9$ boxes, and ``$<10^2$'' counts the number of rules
that touch $10-99$ boxes.  From this, we can see that we have similar
numbers of rules that touch a single box as we have rules that touch
$10-99$ boxes, with some falloff once we consider rules that impact $100$
or more boxes.

For each frequency bin, we see an upward slope from Android~4.3 through to
8.0. Keep in mind that the $y$-axis is log-scaled, so these represent an
exponentially growing number of rules in each bin, which is consistent with
our earlier measurements in Figure~\ref{numrules}.

There is an interesting data point in Figure~\ref{avgnumbyrule}. In Android
9.0, the bin $10000-99999$ has grown from 1 to 5. The four new rules are:
\begin{Verbatim}[fontsize=\footnotesize]
   allow domain dev_type:lnk_file r_file_perms;
   get_prop(domain, core_property_type)
   allow domain fs_type:dir getattr;
   allow domain fs_type:filesystem getattr;
\end{Verbatim}

These four rules allow the attribute
\textit{domain}, which includes virtually every subject, to read all
``link files'' (i.e., symbolic links), to invoke the \texttt{stat()}
system call on virtually every file and directories in the system,
and to access all the ``core properties'' (i.e., configuration
values similar to the registry in Microsoft Windows).  In addition,
even though there are only five rules in this bin, they impact more
than $110,000$ boxes. Therefore, if there is ever a
need to deny the \texttt{stat()} call for a few specific processes,
one would need to carve out additional rules from these generic rules, 
e.g., by creating certain negative rules or splitting these generic rules into 
smaller, more specific ones.

Lastly, we present the ``average'' number of boxes per rule across each
Android security policy snapshot. We see a spike in Android~4.4, 
which is likely caused by the massive  ``unconfined'' domain issue. It then drops for 5.0, and 
becomes relatively stable from 6.0 through 9.0. This suggests that Android
team has achieved a stable engineering discipline, in the sense that while
the absolute number of boxes is growing exponentially, the effort per box
to construct suitable policies is roughly constant across these three
releases.

\vspace{1mm}\noindent\textit{Takeaway \#2:} Some small number of rules
could touch a large number of boxes due to the abstraction
features. Overall, the number of boxes touched by a rule increased over
time, but this ratio seems to be stabilizing, despite the continued
exponential growth in the absolute number of rules and boxes.

\subsection{Number of rules per box} 

\begin{figure}
	\centering
	\includegraphics[width=8.5cm]{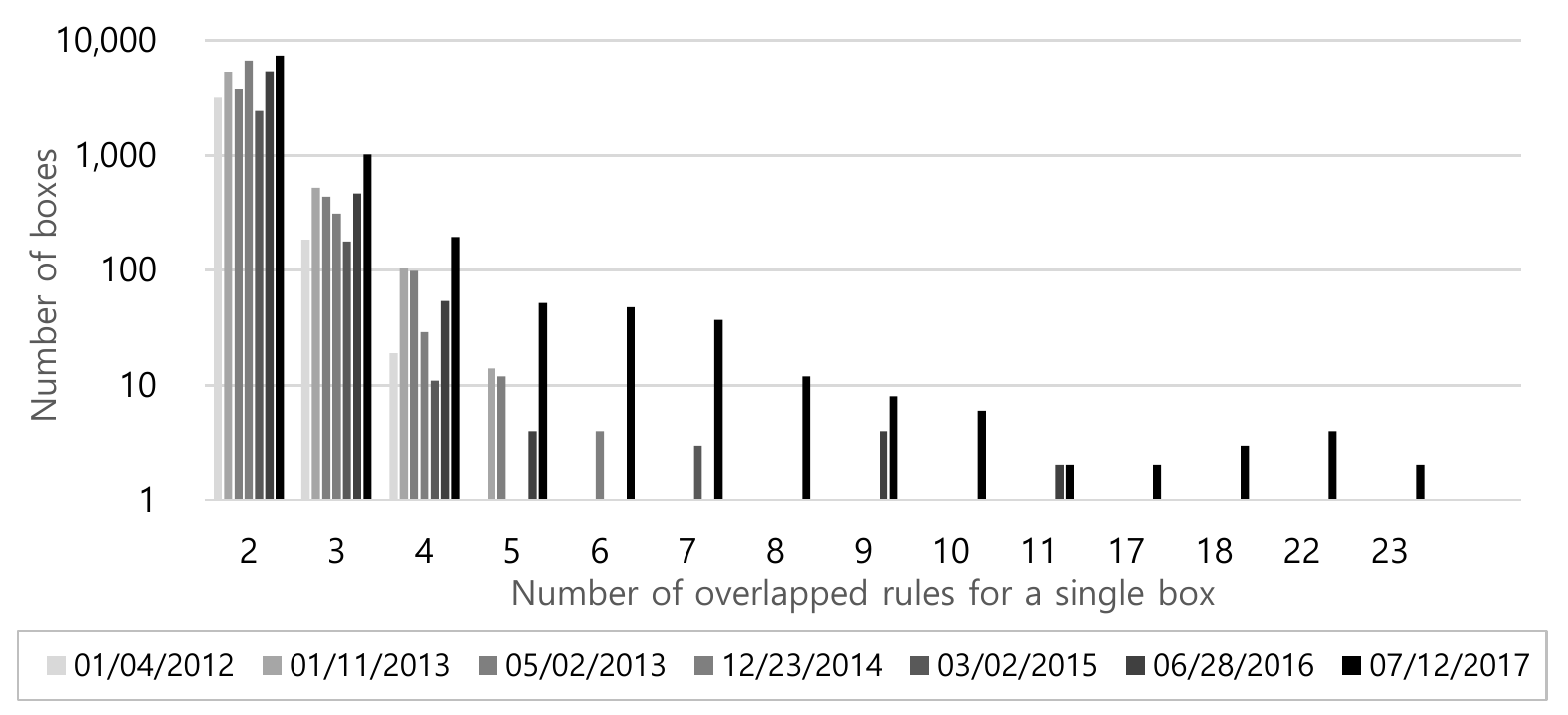}
	\caption{Number of rules mapped to a single box.\label{acsperrule}}
\end{figure}

As before, we define a box to be an atomic access control unit. Our next
concern is that multiple rules might speak about the same box. Why is it
bad to have multiple rules influencing the same box? Consider the case of a
security engineer, auditing the security policy, who determines that one of
the rules is over-broad and thus requires some effort to fix the problem.
If our engineer didn't realize that multiple rules allowed the same box,
then this engineering effort might not have its desired impact.  Where a
high ratio of ``boxes per rule'' indicates the use of macros and grouping
features, a high ratio of ``rules per box'' instead indicates a degree of
imprecision in the design of the rules.

Figure~\ref{acsperrule} shows the rules per box for seven different commits
at dates selected across the range of Android. Since it is natural that one
box should be derived from the only one rule, we do not collect these
cases. As we see from the graph, there are a large number of boxes derived from 2 or 3 rules for all the sampled commits. More importantly, there is a trend that the tail (i.e., boxes with ten or more rules that specify them) is growing over time. The most popular box, generated by a remarkable set of 23 rules is:
\begin{Verbatim}[fontsize=\footnotesize]
   system_server system_file:dir search;
\end{Verbatim}
Some example rules that generate this box include:
\begin{Verbatim}[fontsize=\footnotesize]
   allow hal_gnss system_file dir {open... search};
   allow hal_power system_file dir {open... search};
   allow hal_thermal system_file dir {open... search};
   ...
\end{Verbatim}
HAL, the ``hardware abstraction layer,'' represents an important boundary
between the core Android distribution and the efforts that OEMs make to
port Android to their specific devices.  These 23 HAL-related rules allow
the same box, but it's unclear whether this was deliberate or
accidental. Certainly, if a later analysis determines that {\tt
  system\_server} permissions need to be customized, 
or that this particular box needs to be denied, then each of these 23 rules 
would need to be changed. Many of these rules say nothing at all about 
{\tt system\_server}, at least not directly, but they impact it nonetheless. 

How did this happen? This is a consequence of the overlapping rules in
HAL-related macro functions and attributes, where macros sometimes even expand into other
macros. The HAL subsystem would benefit from some degree of refactoring to
simplify its security policies, redesigning it to avoid so much
overlap.

\vspace{1mm}\noindent\textit{Takeaway \#3:} Most boxes only have a small number of rules, but a few boxes contain more than ten rules, 
which could be an obstacle for effective policy maintenance. 

\subsection{Ratio of rule vs. box changes}

\begin{figure}
	\centering
	\includegraphics[width=8cm]{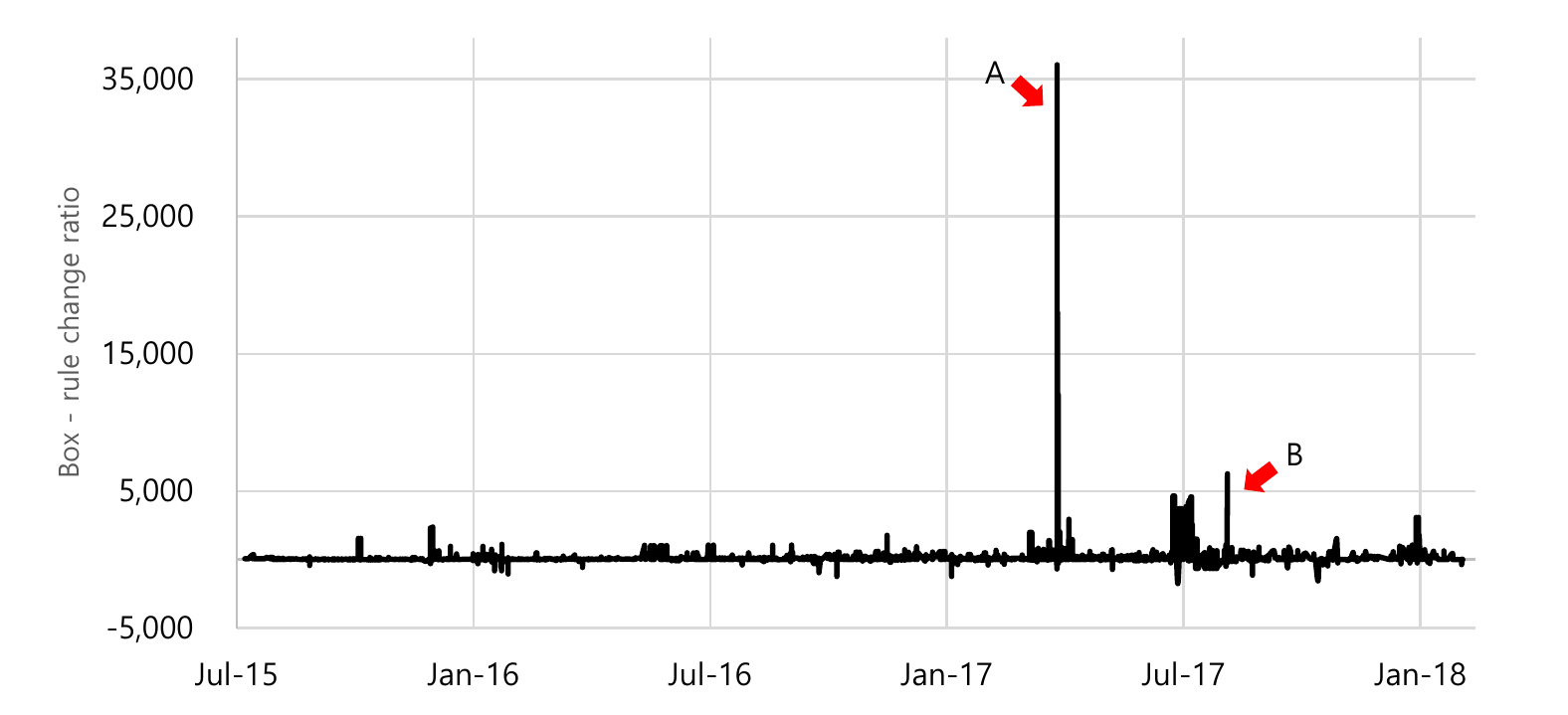}
	\caption{Ratio between change of rules and boxes.}\label{ratio}
\end{figure}

To further understand how commits affect the number of rules and the number of boxes, we plot the ratio $\Delta B/\Delta R$ for each commit, 
where $\Delta B$ is the number of added boxes, and $\Delta R$ is the number
of added rules. If $\Delta B$ (or $\Delta R$) is negative, 
this means that a commit has reduced the number of boxes or rules, respectively.

Figure~\ref{ratio} shows the results starting from July 2015---we did not present earlier data due to the use of \texttt{\small{unconfineddomain}}. 
The average ratio is $120$, meaning that the ``box'' metric is more sensitive to changes. Moreover, we observed negative ratios, which means 
that the number of rules has increased (or decreased) while the number of
boxes has decreased (or increased)---further evidence that their relation 
is non-linear. 

For the peak indicated by A, the ratio is larger than $35,000$. We found that this is because Google added a new file type called 
\texttt{\small{vendor\_file\_type}}, and added file-related rules to the
\texttt{\small{domain}} subject, which includes virtually all processes. 
Another peak, as indicated by B, has not affected the number of rules much; rather, a new permission \texttt{map} is added to existing 
rules. For instance, a rule 
\begin{Verbatim}[fontsize=\footnotesize]
allow domain system_file: file {execute read open};
\end{Verbatim}
would become the following: 
\begin{Verbatim}[fontsize=\footnotesize]
allow domain system_file: file {execute read open map};
\end{Verbatim}

\vspace{1mm}\noindent\textit{Takeaway \#4:} A change in the number of rules
may not always lead to a corresponding change in the number of boxes. 

\subsection{Summary} 

To summarize, the box and rule metrics have an interesting and non-linear
relationship; used in combination, anomalies and large changes directly
point to interesting and relevant engineering changes in Android's history,
such as the introduction and eventual elimination of Android's use of {\tt
  unconfineddomain}.

By expanding rules to boxes, we gain an instrument that is very
sensitive to policy changes. Even though the written rule changes
might be small, the box changes can be enormous. This allows us to
focus our attention on both local discontinuities, which point to
specific significant patches made to the SEAndroid policy, as well as
the broader multi-year trends in SEAndroid engineering. 
With our work, Google and OEMs could institute {\em
security policy metrics} for SEAndroid. Our policy metrics could
prove useful alongside other traditional software engineering metrics (e.g.,
lines of code, or bugs filed and fixed) to help Android project
managers quantify the evolving complexity of their system. Any
non-trivial changes in the metrics might imply significant policy differences, 
or perhaps even inadvertent policy misconfigurations. 
\section{An Historical Analysis}
\label{sec:historical} 

As we mentioned before, the metrics we've devised allow us to plot
long-term patterns of the evolution of the SEAndroid security policy. With
these metrics, and the attendant timestamps, we can examine a number of
other trends over time.

\subsection{The ``age'' of rules}
\begin{figure}
  \centering
\subfigure[Subjects]{\label{sbj_cdf}\includegraphics[width=36mm]{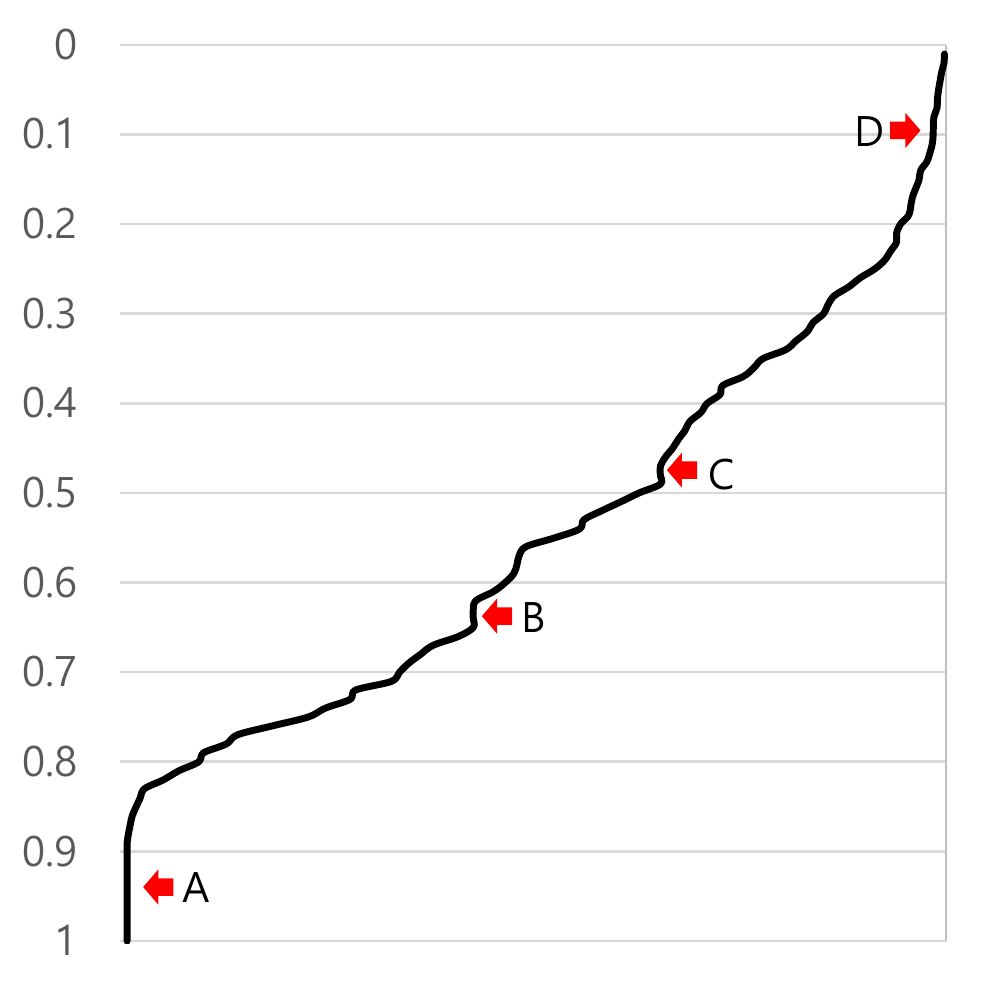}}
\subfigure[Objects]{\label{obj_cdf}\includegraphics[width=36mm]{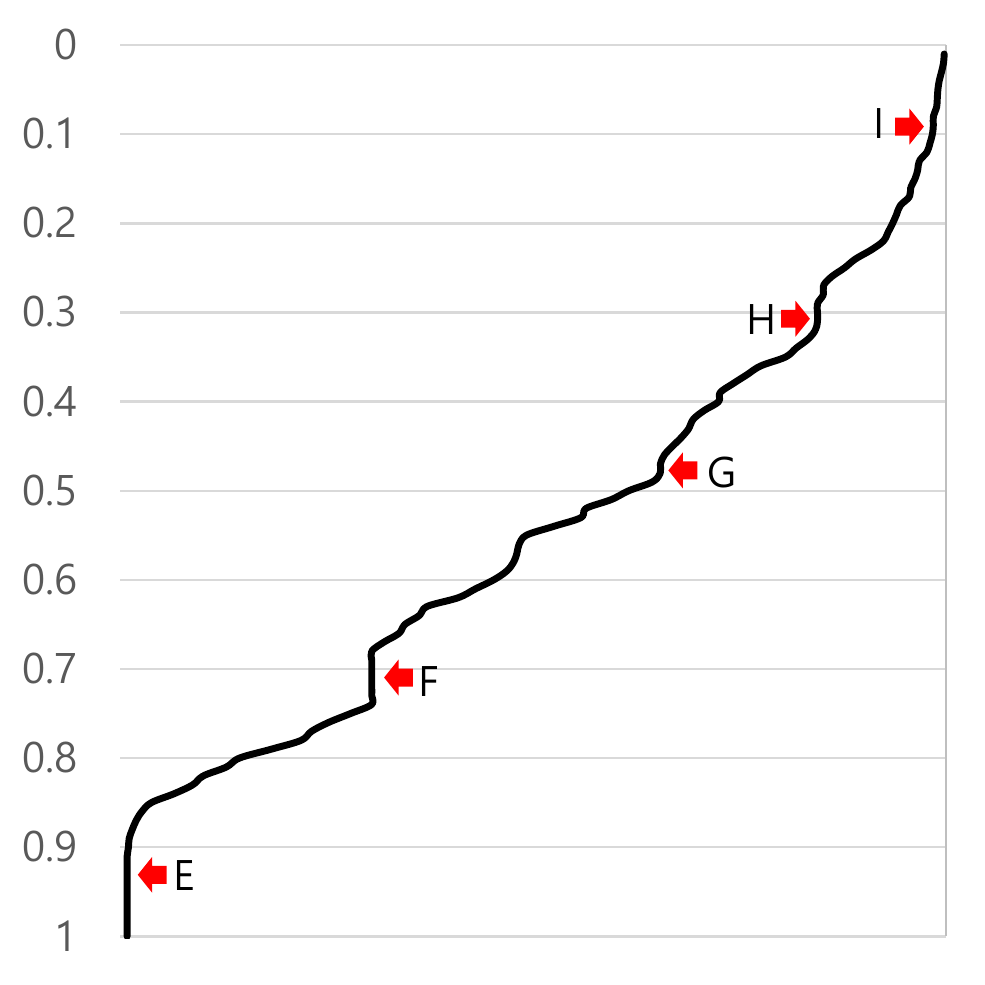}}
\caption{Cumulative distribution of the age of types.\label{typeage_cdf}}
\end{figure}

To start, we want a metric that speaks to the degree of turnover of the rules in SELinux policies. We can derive a given rule's ``age'' by identifying the Git commit when it was most recently changed. The longer a rule has been unchanged, the older it must be. Rather than using wall-clock timestamps, we instead use the order of the commits. This means that, as the rate of commits has increased in recent years, our age metric will ``speed up'' to reflect the increased degree of activity.

Figures~\ref{sbj_cdf} and \ref{obj_cdf} show the cumulative distribution function diagrams of the age of each SEAndroid subject and object, respectively, 
across all commits on the master AOSP branch. The $x$-axis is the age of
the type, and the $y$-axis is the cumulative distribution. 
This means that the types on the left side of the CDF are older, and types
on the right side are more recent.
The arrows ``A'' and ``E'' indicate the ``oldest'' subjects and objectives, respectively, which were created at the very beginning of the policy by NSA; 
they account for $20\%$ of all the current types. 
The arrows ``C'' and ``G'' indicate the introduction of the types for the hardware abstraction layer (``HAL'') subsystem (e.g., \textit{hal\_audio\_ default} and \textit{hal\_drm\_default}). These types were introduced in a single commit by Google in October 2016.

At arrow ``B'', the types of system properties were separated into multiple types such as \textit{audio\_prop} and \textit{bluetooth\_prop}. Before this commit, all the properties were associated with a single type, \textit{system\_prop}. This was presumably part of a privilege separation engineering push within Android. Perhaps as part of the same effort, at arrow ``F'', all the system services were separated into multiple types such as \textit{alarm\_service} and \textit{cpuinfo\_service}. Before this commit, all the system services were assigned to a single type, \textit{system\_service}. In this commit, a new attribute \textit{temp\_system\_server\_service} was added and all the separated services were associated with the new attribute; they inherited all the rules for the existing {\em system\_service} type. The \textit{temp\_system\_server\_service} attribute later disappeared, perhaps unsurprisingly, 
given it is a ``temporary'' name.

At arrow ``H'', a number of HAL related types were added in a single commit. In this case, the new types were related to a new HAL service which is now part of the version 8.0. Regions near arrows ``D'' and ``I'' indicate a radical increase of the types for Android, which are all part of the version 8.0 or later such as lowPAN (low-power wireless personal area network) for Android IoT devices and exported property feature for debugging annotations. 

All of the commits at these labeled arrows after the initial NSA version
were authored by Google. This suggests that other vendors are making only
minor updates and tweaks, at least as far as their contributions to Android
AOSP are concerned; large-scale engineering shifts are only happening at
Google. We discuss more about different vendor contributions in
Section~\ref{sec:contributors}.

\vspace{1mm}\noindent\textit{Takeaway \#5:} A non-trivial portion of the rules were present since the beginning, although the majority of the rules 
have gone through changes. Major events tend to affect a large number of rules, showing up as jump points in the curve. 

\subsection{The increasing policy complexity}

As shown in Figures~\ref{numrules} and \ref{typeage_cdf}, the complexity of the policy is exponentially increasing over time. 
Since the complexity is directly related to its maintainability, we would like to evaluate and estimate the growth pattern in more detail.

\begin{figure}
	\centering
	\subfigure[Types]{\label{fit_type}\includegraphics[width=27mm]{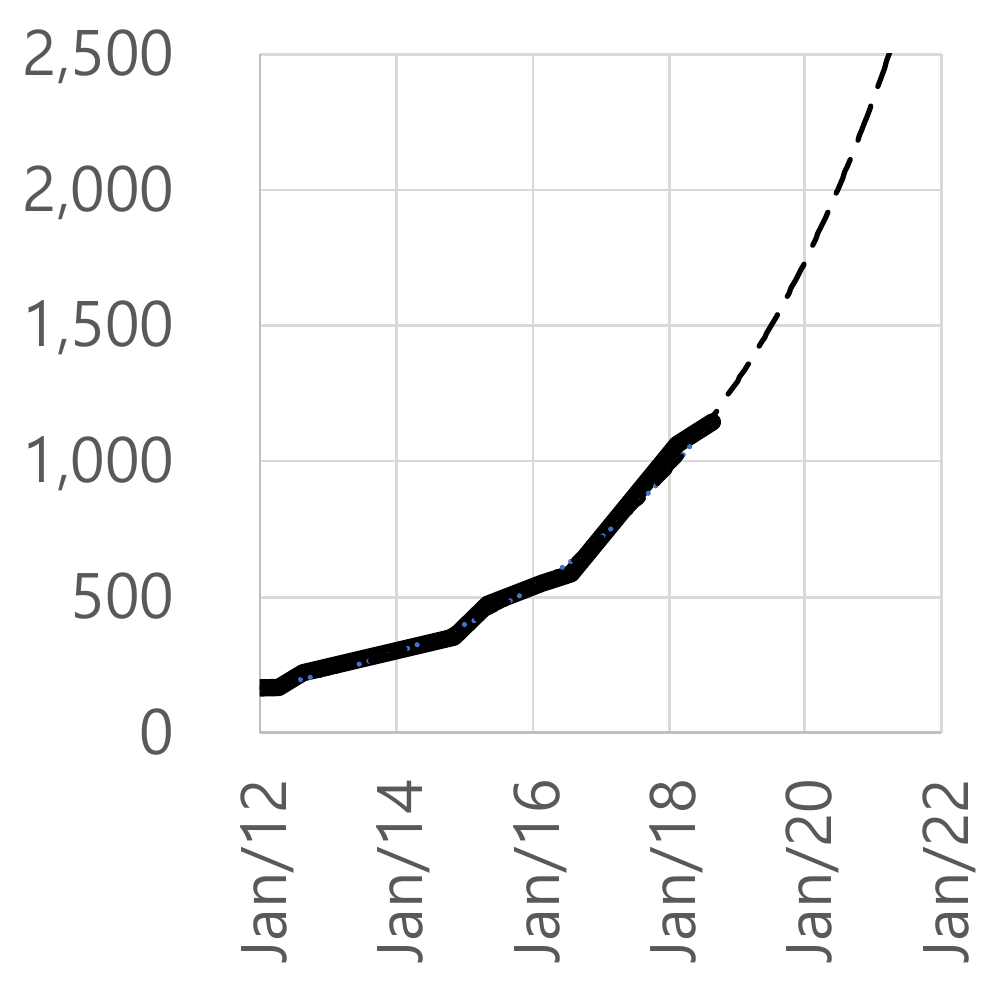}}
	\subfigure[Rules]{\label{fit_rule}\includegraphics[width=27mm]{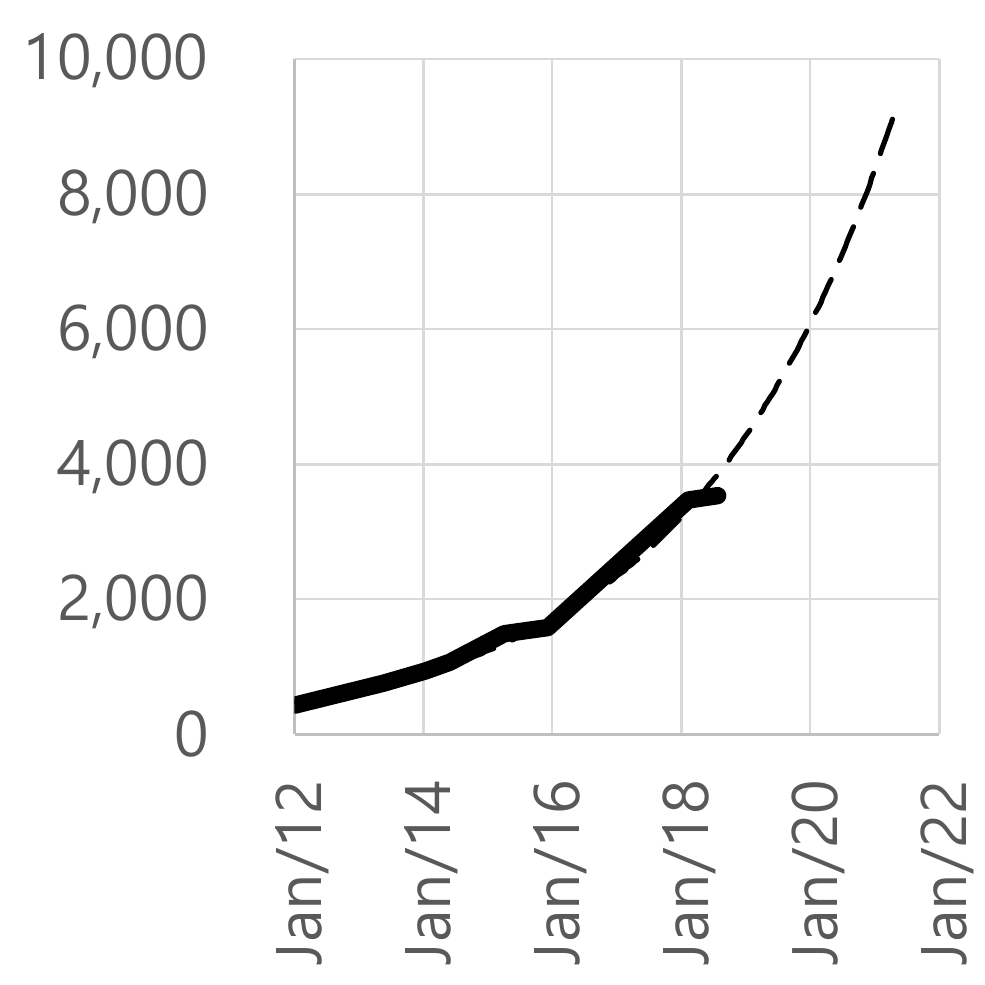}}
	\subfigure[Boxes]{\label{fit_box}\includegraphics[width=27mm]{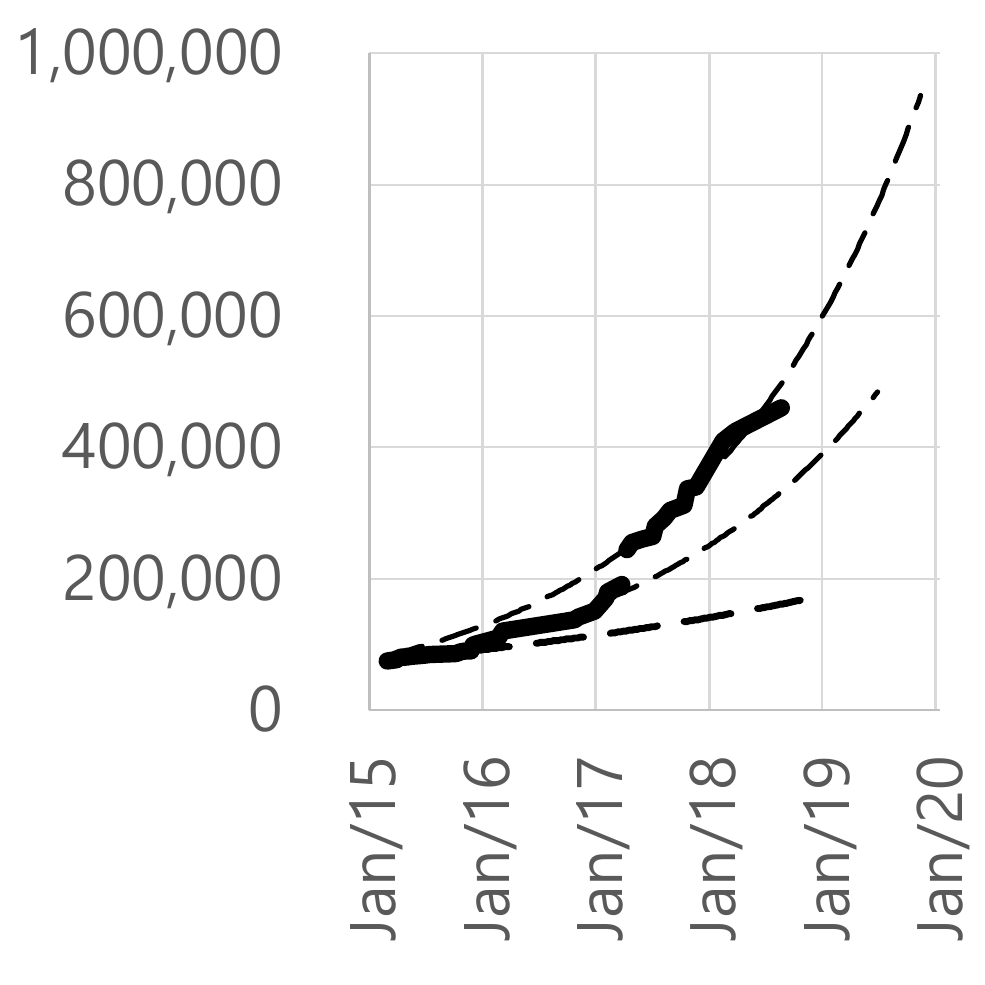}}
	\caption{Fitted curves for types, rules, and boxes.\label{fitline}}
\end{figure}

Figure~\ref{fitline} shows the linear scale fitted curves of three important indicators for policy complexity, 
which are the number of types, rules, and boxes.
We generated this figure partly by extrapolating the future complexity
increase based on the current change trends. 
For instance, extrapolating from the growth, the number of types and the
number of rules might double from today (mid-2018) to mid-2020.

However, the fitted curve of the number of the boxes is different. As shown
in Figure~\ref{fit_box}, we fit different segments of the graph with three 
different curves having different exponents. Not only is the number of
boxes growing, the exponent of growth is growing! Super-exponential growth
of our security policies cannot be a desirable attribute.

In Figure~\ref{fit_box}, we sampled important points after the
``\texttt{unconfined domain}'' period, which is separated to 3 different
curves with different exponents and constants, because there are multiple
jumps in the original curve due to the addition of new types, rules, and
macros in every version release. For example, a big jump happened in 2017,
due to the addition of HAL layer, as shown in Figure~\ref{typeage_cdf}. The
slopes of the fitted lines of all those three different curves are
radically increasing. 

This is a very important indication of the complexity increase---the exponent seems to be increasing, which suggests that the policy maintainability is getting more challenging as time goes on. Since there is virtually no maintenance manual or any public documentation about the policy, the device manufacturers, who need to customize the policy for their products, would face an enormous obstacle in terms of maintainability and evaluability for their product releases.
Even if major contributors such as Google have consistently maintained internal documents, the trend still indicates that their 
maintainability is likely to get more difficult over time. 

\vspace{1mm}\noindent\textit{Takeaway \#6:} The complexity of the SEAndroid
policy is increasingly super-exponentially, which complicates maintainability and analysis.

\subsection{The effect of multiple branches}

\begin{figure}[t]
	\centering
	\subfigure[Number of types captured in May  2017]{\label{num_type_nougat}\includegraphics[width=90mm]{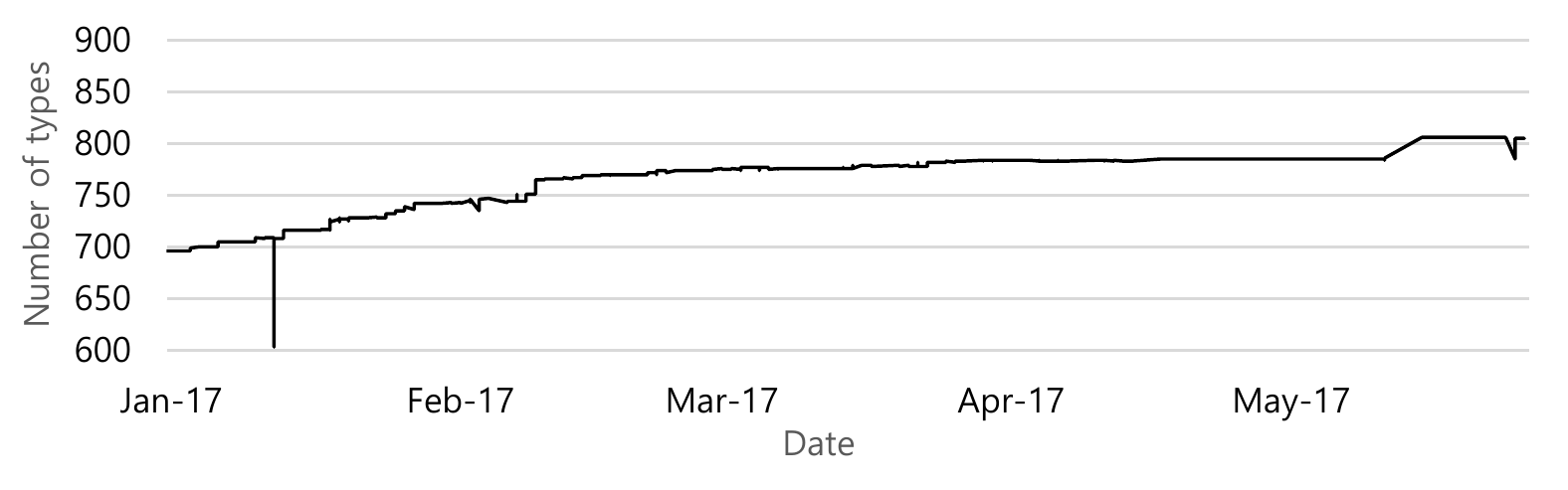}}
	\subfigure[Number of types captured in September  2017]{\label{num_type_oreo}\includegraphics[width=90mm]{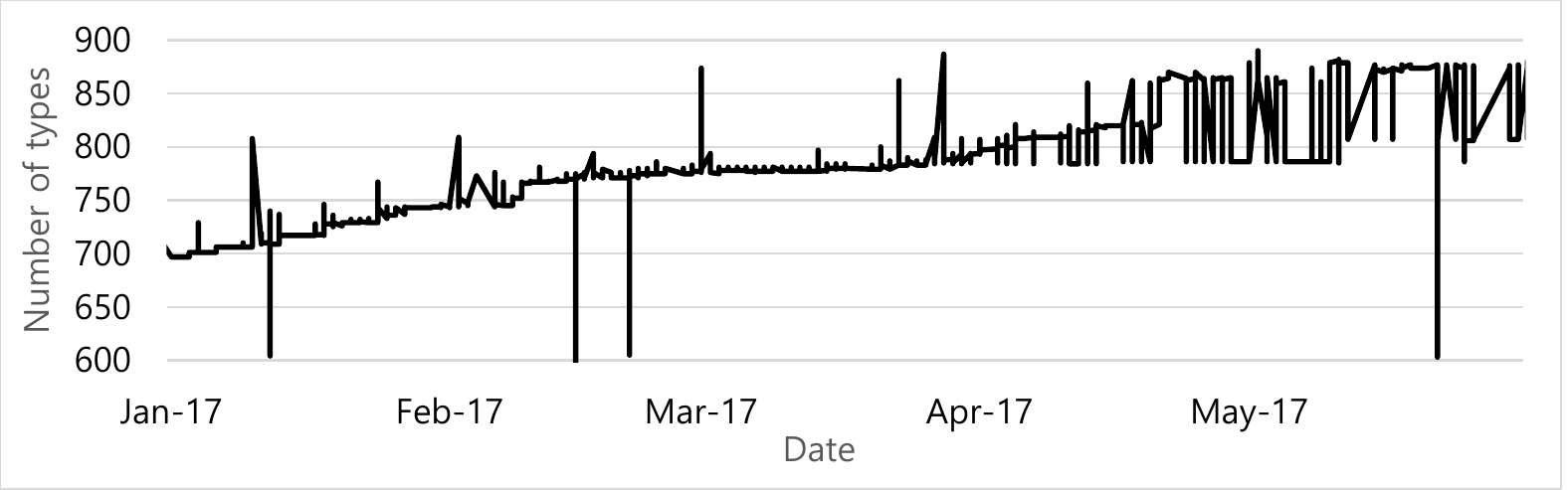}}
	\caption{Comparison of the number of types in same period of time captured in different time}
	\label{num_time_diff}
\end{figure}

Git is a distributed source-code management system, allowing separate
organizations to maintain separate repositories with separate histories,
merging those histories whenever they want. This has the curious property
that the version of history that we see will change over time as
alternative histories are merged or rebased into the mainline history.
Among other issues, if two different branches are merged into one, we
cannot distinguish which was the master and which was the merged branch
without prior information; instead, we only can see that two branches
merged. This means that tracking the ``real'' history of Android AOSP would
require us to maintain snapshots of the entire Git repository's state,
taken over time, rather than just examining the newest repository for its
historical commits; 
this issue has been pointed out before by Bird et al.~\cite{bird2009promises} as one of
the potential downsides of Git mining and analysis.  However, capturing
\textit{all} Git histories, continuously, is infeasible.
To at least understand the impact of these issues, we analyzed our snapshot
of the Android repository before and after the ``Oreo'' Android releases in
August 2017, where we observed more than 1,500
commits added in the span of a few days.

Figure~\ref{num_time_diff} shows the two different curves with the number
of SEAndroid types for the same period of time from January 2017 to
May 2017, but
reflecting the ``pre-Oreo'' and ``post-Oreo'' repository states.
In Figure~\ref{num_type_nougat}, we plot the ``pre-Oreo'' (late ``Nougat'')
data from the end May 2017; in Figure~\ref{num_type_oreo}, we plot the
same time period from the AOSP repository as of September 2017.
These two curves show
the same upward trends but Figure~\ref{num_type_oreo} shows significantly
more noise. This is the result of a merge in the
repository, performed at the end of July, with the merged commits
originally committed in early 2017; these new commits only appeared on the master
branch after the ``Oreo'' release, despite predating it.
Since the graph uses the commit time as its $x$-axis, it shows fluctuations between the
pre- and post-Oreo branch merger. Needless to say, branch merge events
add additional complexity and noise to our data. 

Most of the time-series graphs we derived in this paper 
have a similar square-shaped noise, likely due to merged
branches. Because we cannot track private branches prior to their
merger, it is not possible to distinguish which
stream of the commit was the mainstream of the master branch before
the merge. The timestamps are all we have, and thus we're stuck
looking at interleaved time-series data, and thus the square-shaped
noise. We can at least visually interpret the tops and bottoms of the
square-shapes as representing the two original pre-merger commit
streams. 

This pattern shows that Google is continuing to develop Android
internally, with mergers into the publicly visible AOSP only happening
on major releases. While Android OEMs will hopefully have a more
granular view of Google's efforts, any external developer trying to
work with the SEAndroid security policy will observe these giant
shifts and will have to debug and merge their changes with every
Android release. Even for an Android OEM, Google's large changes
presumably make it difficult for vendors to deviate from AOSP's
SEAndroid policies in any
meaningful way, as they would still have to rebase and/or merge their
changes.  For example, as we can see from the arrows in Figure~\ref{typeage_cdf}, a
single commit could introduce many new types, which would represent a
significant integration challenge for an OEM trying to maintain a
custom policy.

\vspace{1mm}\noindent\textit{Takeaway \#7:} The existence of multiple Git
branches and merges introduce significant measurement challenges, as well
as significantly impacting any third-party attempt to do their own work in
SEAndroid.

\subsection{Case study: Stagefright}

We now turn to examine a newsworthy security event in Android's history and
see if we can observe evidence in the evolution of the SEAndroid
policy to respond to it.

To select the most suitable security events, we looked at
$234$ high-scoring Android CVEs
in a third-party study~\cite{cvedetail}. $128$ of them are
related to hardware and driver issues, which are 
beyond the scope of SEAndroid. Next up are $62$ vulnerabilities with
the media framework. There are also $22$ Adobe Flash
vulnerabilities, but we cannot observe any artifacts of these
in the SEAndroid policy, as Flash is not part of AOSP.
Therefore, media framework vulnerabilities seem like a good place for a case study. 
The most widely publicized security event in the media framework,
in particular, was the ``Stagefright'' vulnerability. 

Stagefright was found and reported to Google by security
researcher Joshua Drake in April 2015~\cite{stagefright}. Google shared the
issue with device manufacturers in May 2015, and the vulnerability was
publicly disclosed in August of that year. This vulnerability is a simple integer
overflow bug in the {\tt libstageflight} library which is used by the media
framework. The attacker can inject carefully crafted malicious code, in a
media file, via any application which uses Android's media API,
allowing the attacker to have arbitrary malicious code execution in
a context with elevated security privileges relative to any regular
user-installed Android application.
Google patched the integer overflow bug and distributed the patch in May 2015.

\begin{figure}
	\centering
	\includegraphics[width=9cm]{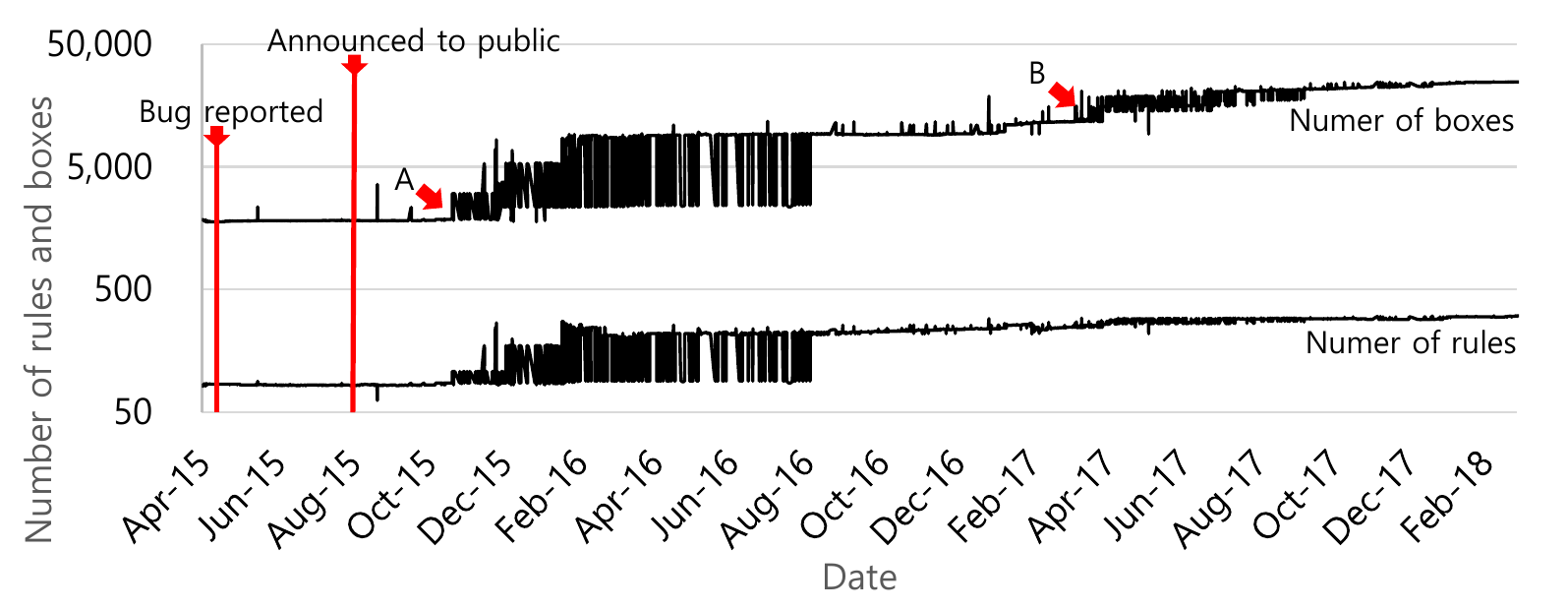}
	\caption{Number of boxes related to media.\label{submedia}}
\end{figure}

Figure~\ref{submedia} shows the number of the boxes and the number of the
rules, in log scale, with media-related subjects and objects from January
2015 to September 2017. Red vertical bars indicate significant events in
the history of the Stagefright vulnerability. The SEAndroid policy did not
change much during the initial period when the vulnerability was
discovered, patched, and the fix was distributed. However, both the number
of boxes and the number of the rules dramatically increased in early 2016,
shown at arrow ``A''. This was a result of separating the media server into
multiple distinct services~\cite{media_hardning} such as
\textit{mediaextractor}, {\em camera\_server}, {\em mediadrm\_server}, and
\textit{mediacodecservice}, each with more limited privileges than the
original monolithic service. Of note, all the media related services lost the
permissions to write a normal file. Additionally, they lost the {\em memexec}
permission.

Another jump, shown at arrow ``B'', is related to the addition of rules
related to a new \textit{audio\_server} service. While the number of rules
changed was small, the number of boxes changed was much larger,
indicating the importance and reach of the changes.

\vspace{1mm}\noindent\textit{Takeaway \#8:} SEAndroid policies are an
essential mechanism for implementing privilege separation, refactoring
monolithic services into smaller cooperating services with more limited
permissions. These policy changes are more visible in our ``box'' metrics
than when just looking at the number of rules.

\subsection{Contributor comparison}
\label{sec:contributors}

\begin{figure}
	\centering
	\includegraphics[width=8.5cm]{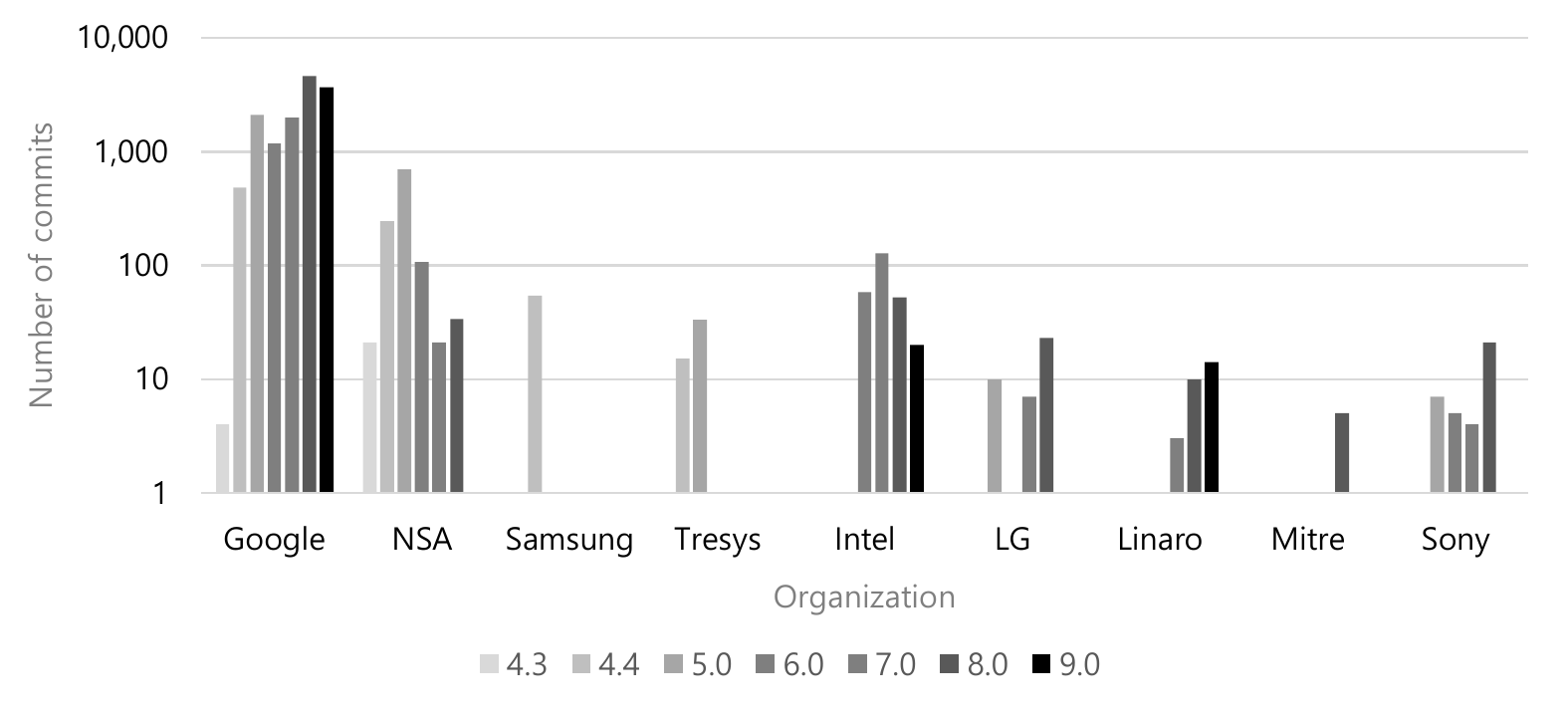}
	\caption{Number of commits contributed by each organization.\label{perorg}}
\end{figure}

Lastly, we perform a measurement to understand the composition of
contributors to the SEAndroid policy.  The Git history associates an email
address with every commit. Ignoring the username and focusing on the domain
name, we can then associate each commit with the author's organization. In
Figure~\ref{perorg}, we show the number of commits applied to each
major Android release, as authored by each organization. As might be
expected, the vast bulk of the commits are authored by Google, and Google's
commit frequency is increasing over time, mirroring the exponentially
growing numbers of rule and boxes.

The NSA is the second most frequent committer, with a spike of commits in
the earlier Lollipop release, indicating some degree of NSA assistance as
Google moved away from the ``unconfined'' domain design toward a more
rigorous security policy.

We also see device and hardware manufacturers (Samsung, Intel, LG,
Sony). Each of these vendors maintains its own private policies that they
ship with their own devices. It is in a manufacturer's interest to
contribute patches back to Google to avoid too much drift between Google's
codebase and their own. Samsung stands out, in this regard, for having zero
visible commits in the latest four versions of Android. (Perhaps Samsung
contributed its policies to Google without sending Git pull requests. If
so, a Google engineer would have merged their changes by hand, and the
resulting commits would appear with a Google email address.)

The remaining organizations, Tresys, Linaro, and Mitre, represent two
commercial consultancies and a Federally Funded Research and Development
Center (FFRDC) that works extensively with the U.S. government. Tresys
notably offers SELinux policy customization as a service for paying
customers~\cite{tresys_selinux}.

\vspace{1mm}\noindent\textit{Takeaway \#9:} Even though there are a number
of non-Google contributors, most of their commits are simply fixing typos
or adding missing simple rules which imply that only Google is leading the
project. In addition, all the important policy changes such as all the
arrow marks in Figures~\ref{numrules} and~\ref{typeage_cdf} are authored
only by Google.
\section{Discussion}
\label{sec:discussion} 

Next, we discuss three items related to our measurement study: the Tizen system, 
the new Android Treble release, and Android for Work. 

\subsection{SEAndroid vs. Smack}

As we have seen, the SEAndroid policy is getting
more complex over time. Sophisticated policies may promise better security,
but they also make it challenging to reason about the configuration,
allowing innocent mistakes to creep into the design.

As a point of contrast, Tizen~\cite{tizen} uses a very different
approach. Just like Android, Tizen is a Linux-based operating system
targeted for mobile devices, and it uses a similar access control mechanism
to SEAndroid called Smack~\cite{schaufler2008simplified}. The biggest
difference is that Tizen doesn't overload Unix user IDs to separate
applications from one another. 
Instead, all the apps for the same user use the same Unix user ID, but with Smack labels to isolate the applications.

\begin{figure}[h!]
	\centering
	\includegraphics[width=8.5cm]{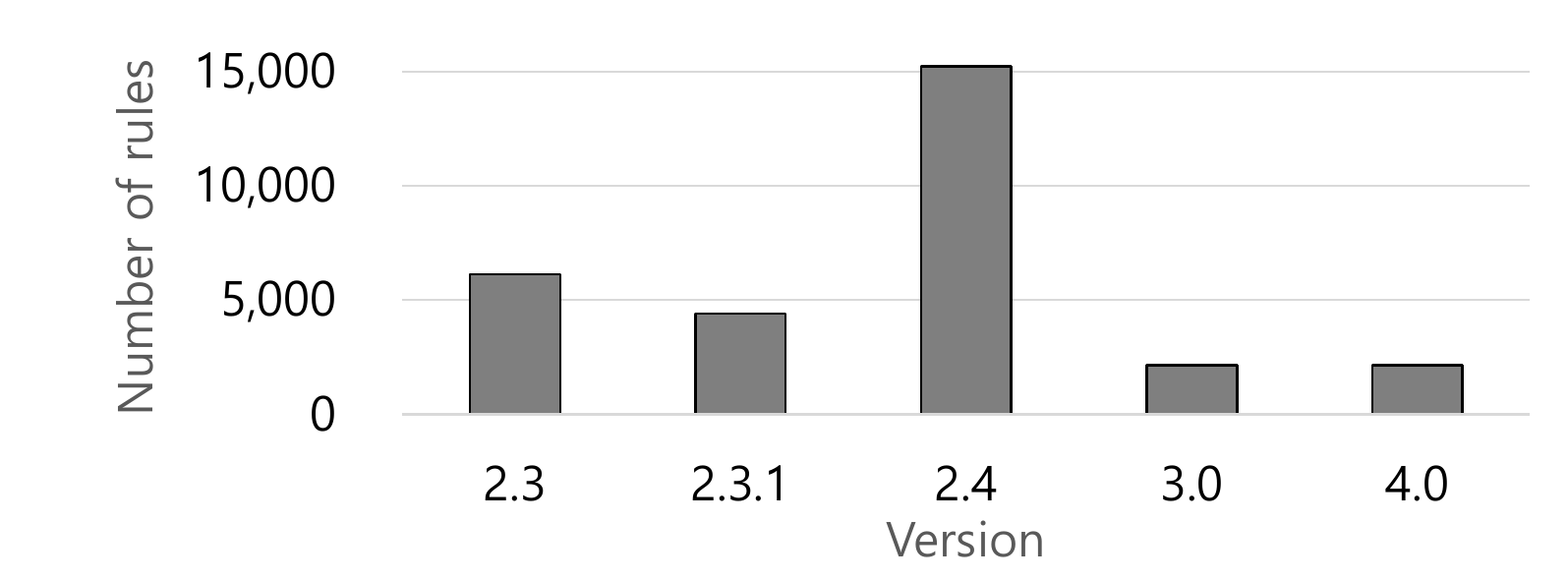}
	\caption{Number of Smack rules in Tizen releases}
	\label{tizen_rules}
\end{figure}

We performed a small-scale experiment by counting the number of Smack rules in Tizen from its Git repository, and 
Figure~\ref{tizen_rules} shows the result. We have removed the Smack rules
related to specific applications to make a fair comparison with SEAndroid. 
We can see that the number of rules fluctuates across versions, but it stabilizes at roughly 2,000 in Tizen 3.0, 
which is radically smaller than the number of rules in SEAndroid.

In Tizen 2.x, Smack is used for all the access control mechanisms including
application isolation and API permissions, with the number of rules tightly
related to the number of features. Consequently, the policy complexity is
relatively high.  The decrease in 2.3.1 and increase in 2.4 in
Figure~\ref{tizen_rules} is simply due to the removal and addition of many OS
features~\cite{tizen_2.x}.  Tizen 3.0~\cite{tizen_3.0} featured a complete redesign
of the Smack policy, providing only a minimal privilege
separation for the system resources and each
application~\cite{tizen_3.0_smack}: Unix user IDs and Linux namespaces are
used to isolate system services, and a new module called 
Cynara~\cite{cynara} was added to manage 
per-app API permissions (i.e., as in Android, users' grants of specific
permissions to Tizen apps is no longer managed by the underlying Smack
system but is instead managed in a separate system).  The number of the rules in Tizen 3.0 and Tizen 4.0 is
exactly the same: 2,134 rules, despite the otherwise significant changes
between the two major releases, including support for IoT devices, and
coding in C\#/.NET~\cite{tizen_4.0}.  

Although there is no ``correct'' approach to access control configuration,
and we take no position on whether Tizen is fundamentally more or less
secure than Android, Tizen demonstrates that MAC policies can be deployed
without the staggering complexity of modern SEAndroid policies. Of course,
the very lack of fine-grained permissions expressed in Tizen could as much
represent the benefits of simplicity as the pitfalls of
over-simplification. Regardless, the fact that both Android and Tizen are
attempting to solve quite similar problems in radically different ways
suggests that there may be lessons to draw from each to the benefit of the
other.

\subsection{Android Treble}

In Android 8.0, Google introduced Treble~\cite{treble} as a
framework to separate platform and manufacturer features.
Treble's main goal is to make it easier for Google to ship updated Android
systems by creating a stable abstraction boundary between vendor features
and the Android distributions from Google. If done properly, we should see
a higher fraction of Android devices running recent releases of Android,
improving security for Android users and simplifying the release
engineering process for Android OEMs.

While a full summary of Treble is beyond the
scope of this paper, Treble does have an impact on SEAndroid.
In prior releases, Android vendors would
start with Google's AOSP SEAndroid policy and make suitable
modifications to support device-specific features, perhaps porting
changes forward from release to release. Treble 
separates ``vendor policies'' from Google's own system policy.
Both policies are separately compiled to 
a new common intermediate language (CIL), and are then combined into
a single policy as part of the boot process. This allows Google to
update its SEAndroid policy without vendor intervention.

The AOSP version of the SEAndroid policy is effectively the same as
before, so it doesn't impact the continuing growth of SEAndroid policy
complexity over time. But, now that vendor policies will be separated out, as more
vendors ship devices with Android~8.0 or later, follow-on research to this
paper will be able to look at vendor-specific firmware images as opposed to AOSP releases, 
and such future work will be able to make interesting comparisons between vendors. Some
vendors will inevitably make huge changes while others change little or nothing.
And, inevitably, some vendors will introduce security flaws by enabling too many permissions.
Also, the new CIL format retains some macro and grouping structures, allowing it to be far more amenable to analysis than the
compiled binary policies of earlier Android releases, where macros have
been completely expanded.

\subsection{Android for Work}

Android for Work is an enterprise security solution for
``Bring-Your-Own-Device'' (BYOD) environment, which was introduced in
Android~5 and is still under active development. Android for Work supports
separated runtime environments between work applications and personal
applications.  This requires extensive resource isolation, access control,
and policy configuration. Knox, a similar BYOD solution introduced by
Samsung~\cite{knox_wp}, also used SELinux policies to support such access
control. Android for Work would certainly be amenable for the same sort of
analysis that we did in this paper, but the source code for it is not made
available as part of AOSP. If it does become available in the future, its
evolution over time would be an interesting subject to study, particularly
as the Android for Work developers must necessarily respond to and
integrate with changes in the larger SEAndroid policy.
\section{Related work}
\label{sec:related}

Our work is most related to two lines of existing work: analyses of
SEAndroid/SELinux policies, and general software-engineering analyses of Git repositories.

\vspace{1mm}\noindent\textbf{SEAndroid/SELinux policy analysis.} 
Understanding software complexity is an important topic in the software
engineering community, and a variety of useful metrics have been proposed
over the years, such as cyclomatic complexity~\cite{mccabe1976complexity}
and Halstead volume~\cite{munson-icse89}.  However, in the context of
SEAndroid/SELinux policy, complexity measurements are much less studied;
the dominant metric is simply the number of rules in the policy source
code~\cite{reshetova2015characterizing}.  Researchers have also used formal
verification on SELinux
policies~\cite{hicks2010logical,jaeger2003policy,sarna2004policy,jaeger2004resolving,archer2003analyzing,sasturkar2006policy},
artificial intelligence, information flow integrity measurement~\cite{jaeger2003analyzing,jaeger2006prima,vijayakumar2012integrity},
and functional tests~\cite{wang2017spoke}. Even machine learning
techniques have been used to analyze policies based on SELinux denial
logs from billions of devices~\cite{wang2015easeandroid}. 

Chen et al. ~\cite{chen2017analysis} study the SEAndroid policy with the goal of identifying potential 
misconfigurations. They combine the SEAndroid mandatory policies with the discretionary policies embedded in the Android file system 
(i.e., Unix permission bits), giving them a more complete look at what is actually allowed or denied in practice.
Both their work and ours present metrics and tools that might be useful in the Android development process.
One important difference, however, is that we focus on quantifying the complexity of a policy snapshot, and on how the complexity 
evolves over time. 

\vspace{1mm}\noindent\textbf{Git mining.}
Bird et al.~\cite{bird2009promises} provide a comprehensive analysis on
the pros and cons of Git mining~\cite{bird2009promises}.  Notably, one of
the challenges they mention is the lack of a mainline repository when dealing
with multiple Git branches. Many other researchers have studied Git
repositories (see,
e.g.,~\cite{storey2014r,vasilescu2015gender,gousios2016work}). 
Negara et al.~\cite{negara2014mining} use Git mining to detect
patterns of code changes~\cite{negara2014mining}, German et
al.~\cite{german2016continuously} and Jiang et al.~\cite{jiang2013will}
analyze Linux kernel repositories' code over time.  In our work, we apply
similar concepts toward the study of SEAndroid's evolution.
\section{Conclusion}
\label{sec:conclusion}

In this paper, we have performed the first historical analysis of the
 SEAndroid policy to understand its evolution over time.  We looked at both
``rules'' as written in the SEAndroid policy and the ``boxes'' those rules
expanded to after processing all the macro and grouping operators.
By plotting these metrics over time, we can observe the exponential growth
in the complexity of SEAndroid policies, which will inevitably hit a brick
wall of engineering complexity and require new and novel approaches to
manage this complexity. Such approaches might lean on techniques
from across many fields of computer science. 
For instance, we could utilize \textit{neverallow} box as a policy ``coverage'' metric.
Neverallow rules are used as assertion tools that policy engineer defines the boxes
which should never be allowed, but as of November 2018, only 3\% of the possible
boxes are covered by allow and neverallow boxes. Therefore, we could increase the
coverage by measuring the current status and enhancing the number of the neverallow
boxes.

\section{Acknowledgment}
We thank Stephen Smalley, Robert Williams and Tomasz Swierczek for valuable advice
and answering many of our questions as well as the anonymous referees for their valuable feedback.
This work was supported in part by NSF
grants CNS-1801884, CNS-1409401, and CNS-1314492.

\bibliographystyle{abbrv}
\bibliography{historical_analysis}

\end{document}